# Far-field subwavelength resolution imaging by spatial spectrum sampling

*Tie-Jun Huang, Li-Zheng Yin, Ya Shuang, Jiang-Yu Liu, Yunhua Tan, and Pu-Kun Liu*[*]

State Key Laboratory of Advanced Optical Communication Systems and Networks, Department of Electronics, Peking University, Beijing, 100871, China

*Correspondence to pkliu@pku.edu.cn

**Abstract:** Imaging below the diffraction limit is always a public interest because of the restricted resolution of conventional imaging systems. To beat the limit, evanescent harmonics decaying in space must participate in the imaging process. Here, we introduce the method of spatial spectrum sampling, a novel far-field superresolution imaging method for microwave and terahertz regime. Strong dispersion and momentum conservation allow the spoof surface plasmon polaritons (SSP) structure to become a sensitive probe for spatial harmonics. This enables that the spatial information of the targets including both propagating and evanescent components, can be extracted by tuning and recording SSP in the far field. Then, the subwavelength resolution is constructed by the inversed Fourier transform of the sampled spatial spectrum. Using the modified subwavelength metallic grating as the probe, a far-field resolution of 0.17 $\lambda$ is numerically and experimentally verified, and two-dimensional imaging ability is also fully discussed. The imaging ability and flexibility can be further optimizing the SSP structures. We are confident that our working mechanism will have great potentials in the superresolution imaging applications in the microwave and terahertz frequency range.

## 1. Introduction

Achieving spatial resolution without limited by the working wavelength always attracts tremendous attention, due to the pervasive applications of imaging. However, the diffraction limit indicates the fine features smaller than half a wavelength are carried by the evanescent harmonics [1], whose amplitude exponentially decays with distance. The contribution of these harmonics is negligible when the imaging distance larger than a wavelength. To circumvent the inherent limitation, the optical elements should own ability of capturing the evanescent part of spatial information. By mechanically canning a sensitive probe, the near-field microscope (NFSM) [2-4] can well surpass the diffraction limit for thousand times, but the imaging operation is very time-wasting. To obtain superresolution images in a single shot, the concept of perfect lens is theoretically exploited by Pendry [1], utilizing the plasmonic effect in materials to enhance the evanescent waves [5, 6]. Soon after this imaging method is experimentally verified at optical wavelength by plasmonic slabs, such as the noble metal [7, 8] and dielectrics [9]. Although the imaging ability of such lenses can be further enhanced by other methods [10], the working distance is still subject to the near field [11, 12]. A remarkable device, termed hyperlens, is capable of converting the evanescent harmonics into propagating waves through curved geometry [13-16] or gratings [17, 18]. In this way, superresolution pictures are formed in the far field. The hyperlens can be constructed by periodic arrangement of plasmonics materials and common dielectrics [11-20].

When it comes to the longer wavelength, for example, the terahertz domain, which has great potential in non-destructive testing and manufacturing quality control, the metal behaves as perfect electric conductor [21, 22]. To mimic the intriguing properties of natural plasmonic effect, by texturing the geometrical features of high-conductivity metal, the concept of spoof surface plasmons (SSP) [23] is developed with low loss and high flexibility. Based on SSP, tremendous potentials have been corroborated in the microwave and terahertz frequency ranges [24-37], including wavelength waveguides [24, 25], integrated photonics circuits [26, 27], superfocusing waves [28, 29], trapping rainbow [30], controlling waveguide modes [31], ultra-sensitive sensors [32, 33], on-chip terahertz sources [34], and so on. However, the corresponding researches of superresolution imaging based on SSP are rarely reported. Unlike the natural plasmonic effect to offer the enhancement of broadband evanescent fields [7], under certain parameters, the SSP behaves as a narrowband spatial filter. Using SSP structure, only limited spatial information will be captured and participated into the imaging process. Recently, the approach to enlarge the bandwidth of imaging spatial spectrum employs a serious auxiliary sources, which amplifies the spatial harmonics by serval slabs made by different hyperbolic metamaterials [38-40]. But replacing different slabs is still inconvenient, and it is only suitable for optical wavelength. So far, the practical superresolution imaging methods at microwave and terahertz wavelength are still highly desirable.

In this work, we propose the concept of spatial spectrum sampling, a method for retrieving the broadband evanescent information of targets, and demonstrate that it allows almost real-time subwavelength resolution imaging at far field for the terahertz and microwave domain. Our mechanism resides in the momentum conservation between the spatial harmonics of targets and strong dispersive plasmonic modes supported by adjacent devices. As a specified porotype, the modified SSP structure is theoretically, numerically and experimentally studied for the process of sampling, sending and recovering the broadband spatial spectrum in the far field. Then, the subwavelength image is constructed by the Inverse Fourier Transform (IFT) of the spatial spectrum. A resolution of 0.17 $\lambda$ is numerically and experimentally verified. The resolution potentially and flexibility of the imaging process can be further enhanced via optimizing SSP structure. The extension of two-dimension imaging is also numerical obtained by simply rotating the imaging device.

**2. The principle of the imaging method**

The basic principle of our imaging method is illustrated in Fig. 1. In the conventional raster-scanning imaging, the sampling process is realized by mechanically moving a probe point-by-point, as shown in Fig. 1(a). The superresolution images can be obtained by combining the locations with their sampled field values. In contrast to that, as depicted in Fig. 1(b), our ambition is to perform sampling operations in the domain of spatial spectrum $F(k_x)$ which is the Fourier Transform (FT) of targets field $E(x)$.

$$F(k_x) = \int_{-\infty}^{\infty} E(x) e^{-jk_x x} dx, \tag{1}$$

where $k_x$ is the transverse wavevector. Interestingly, for a certain field of the targets, its spatial spectrum is independent on the working frequency $f$. Higher working frequency corresponds to larger wavevector $k_0$, ($k_0 = 2\pi f/c$, $c$ is the speed of light.), leading to a broadband propagating spectrum windows $|k_x|<k_0$. Therefore, a better resolution can be expected at higher frequency.

This sampling process is based on the momentum conservation, meaning that the spatial components of the targets can be extracted by propagating modes of adjacent sampling devices when they have the same transverse wavevectors. In this way, by tuning parallel wavevector of the sampling devices, the spatial profile of the targets is gradually captured. If the evanescent harmonics can be reached and retrieved, the resolution of the recovered images will surpass the diffraction limit. Therefore, having response for the broadband spatial spectrum is the essential demand for the sampling devices. The tuning and sampling processes must avoid the time-consuming mechanical movement, then the imaging speed will be significantly enhanced. With the assist of IFT, the sampled spatial spectrum (shown in Fig. 1(c)) can be translated to the real image. As plotted in Fig. 1(d), two spikes with a center to center distance λ/4 (λ is wavelength of the working frequency) is successfully distinguished by this method.

As a realistic verification, the SSP structure is employed in the imaging process, and the reasons are presented as follows. Firstly, the strongly dispersive and slow-wave properties allow SSP become a sensitive probe for even deep evanescent harmonics; Secondly, by changing the geometry, its working frequency and dispersive features can be designed as desired; Thirdly, the fabricated demands and loss issues of the SSP structure are superior to others plasmonic devices at microwave and terahertz wavelength [41]; Lastly, the working parallel wavevector of the SSP mode is easily to be tuned, such as changing the working frequency [24], changing properties of loaded materials [42], integrating with tunable components [43] and so on.

## 3. Sampling spatial spectrum by SSP

The chosen probe is the subwavelength metallic grating, which can be replaced by any available SSP structures. The schematic of this grating is rendered in Fig. 2(a). The period $p$, width $d$ and height of the grooves $h$ are utilized to describe the geometrical characters. The modal expansion method (MEM) [44, 45] is used to theoretically derive the dispersion relationship under the condition of perfect electric conductor approximation. For simplification, the thickness of the grating in the y direction is supposed to be infinite. After matching the tangential components of electric and magnetic field at the air-grating interface and the bottom of grating, the dispersive equation describing the parallel wavevector $k_s$ of SSP can be obtained.

$$\sum_{-\infty}^{\infty} \frac{\text{sinc}^2\left(\frac{k_{sn}}{2}d\right)}{\sqrt{k_{sn}^2 - k_0^2}} = \frac{p}{d}\frac{\cot(k_0 h)}{k_0}, \tag{2}$$

with $k_{sn} = k_s + 2\pi n/p$, $n$ represents the diffraction order. In Fig. 2(b), we calculate and plot the dispersion curve, under the parameters of $p = 100$ μm, $d = 50$ μm, and $h = 260$ μm. The x coordinate is normalized by the free space wavevector $k_c$ at 0.26 THz, near the cutoff frequency of the foundational SSP mode. It is clear that, the dispersive curve significantly departs from the light line. The parallel wavevector range of SSP is only dependent on the period of grating following the relation $-\pi/p \leq k_s \leq \pi/p$. Under the present parameters, the maximum $k_s$ remarkably reaches up to $5.8k_c$, located at the deeply evanescent domain. The value of maximum $k_s$ can be further improved by adjusting $p$. To visualize the SSP with various $k_s$, the finite element method (FEM) is utilized to calculate the evolution of electromagnetic waves. The material of the grating is set as copper, its conductivity being 5.959×107 S/m. We plot x-component electric field maps in the inset of Fig. 2(b), respectively corresponding to $k_s = -0.8k_c$ (0.171 THz), $k_s = -2k_c$ (0.237 THz), $k_s = 5k_c$ (0.26 THz). From these field distributions, we can see that SSP is evanescent in the vertical direction, and its transverse wavevector can be freely tuned by simply changing frequency due to the strong dispersion.

For a single working frequency point, the parallel momentum of SSP owns two certain opposite values which are related with the backward and forward waves with respect to the x-axis. If the scattering wave impinges on the grating, as shown in Fig. 2(a), by tuning the working frequency from zero to the cutoff frequency, the SSP modes with $|k_s| < \pi/p$ can be sequentially launched without any separation. The amplitude of the launched SSP mode is strongly related to the corresponding spectral component of the incident waves. Note that, due to the flexibility of SSP, its parallel wavevector can also be tuned by other manners to avoid the variation of working frequency, more details of which are discussed in the last section. To give a quantitative demonstration, we record the intensity of the excited SSP from 0.16 THz to 0.23 THz when the parallel wavevector of the incident waves are -2$k_c$, 3$k_c$, -5$k_c$. The simulated results are plotted in Fig. 2(c). Obviously, in the intensity profile of SSP, there are serval sharp peaks excellently coinciding with the input spectral components. The physical mechanisms behind this is easy to be explained by the momentum conservation. We also present a theoretical deduction for the coupling. A TM-polarized incident wave with particular parallel wavevector $k_x$ is expressed by

$$E^i(\mathbf{r}) = \frac{1}{\sqrt{d}} e^{ik_x x} e^{ik_z x} \left( \mathbf{x} - \frac{k_x}{k_z} \mathbf{z} \right), \quad (3)$$

where $k_z = \sqrt{k_0^2 - k_x^2}$. The reflected waves from the grating associated with n-diffraction order can be written as

$$E^{ref,n}(\mathbf{r}) = \rho_n \frac{1}{\sqrt{d}} e^{ik_{xn} x} e^{-ik_{zn} z} \left( \mathbf{x} - \frac{k_{xn}}{k_{zn}} \mathbf{z} \right), \quad (4)$$

where $k_{xn} = k_x + 2\pi n / d$, $k_{zn} = \sqrt{k_0^2 - k_{xn}^2}$ and $\rho_n$ is the n-order reflection coefficient. After considering the boundary condition, the reflection coefficients can be easily extracted

$$\rho_n = -\delta_{0,n} - \frac{2i \tan(k_0 h) S_n S_0^*(1+\delta_{0,n})}{1 - i \tan(k_0 h) \sum_i S_i^2 \frac{k_0}{k_{zi}}}, \quad (5)$$

in which $S_n = \sqrt{\frac{a}{d}} \operatorname{sinc}\left(\frac{k_x^{(n)} a}{2}\right)$. Here, we only calculate the zero-order reflection coefficient, and combine the equation (2) to substitute $k_0$ for $k_s$. The amplitude of theoretical results is plotted in Fig. 2(d). Clearly, the trend of theoretical prediction and numerical calculation matches well. The amplitude of excited SPP appears strongly sensitive to the condition of momentum match, i.e., $k_x = k_s$, which indicates the SSP structure is capable of sampling the spatial spectrum of the near-field scattering waves. The intensity of excited SSP waves is highly dependent on the corresponding spectral component of the incident waves. Therefore, the whole spatial spectrum of the targets can be sampled and reconstructed by adjusting the wavevector of the SSP mode, without any mechanically moving parts.

Considering the practical usage, we always want to perform the imaging operations in the far field. Due to the momentum mismatch between free space waves and SSP mode [46, 47], the extracted information carried by SSP is still stuck in the grating surface. Fortunately, in our previous investigations [48], an ultra-broadband SSP coupler has been proposed and demonstrated to alleviate the momentum mismatch. This coupler can be utilized to efficiently radiate the sampled spatial spectrum into the far field. The structure of this coupler is rendered in Fig. 3(a), where all the parameters are clearly defined. The coupler is composed three parts: a tapered cover plate used for gradually compressing the energy of incoming SSP; a grade grating used for mitigating the momentum mismatch between SSP and TEM mode inside the parallel-plate waveguide (PPWG); a tapered PPWG used for sending the EM energy into the far-field, similar with an antenna. The working performance of this coupler is studied and presented in Fig. 3(b), which is also calculated by the FEM method. The parameters are set as: $L_1 = 2.25$ mm, $L_2 = 2.75$ mm, $L_3 = 2.5$ mm, $L_4 = 1.25$ mm, $L_5 = 0.41$ mm, $N = 9$, $\theta_1 = 14.6°$, $\theta_2 = 14.9°$. Within the frequency band from 0.14 to 0.26 THz, the coupler holds extremely low reflection coefficient, indicating the high-efficiency conversion between SSP and far-field radiation, as shown by the S11. Here, the uniform grating side is defined as the port 1. The field evolution of transmission between SSP and free space waves is depicted in the inset of Fig. 3(b), with the frequency being 0.237 THz. From this field map, we can recognize that the SPP gradually converts to the TEM modes, and radiates into the space through the coupler.

**4. Construct superresolution image**

By combing all the SSP parts, the imaging device is presented in Fig. 3(c). The uniform metallic grating at center is utilized for sampling the spatial spectral information of scattering waves from the targets. The SSP waves carrying spatial harmonics $k_s > 0$ ($k_s < 0$) will propagating toward the +x (-x) direction. Two couplers at both ends can send the sampled information into the far field. By collecting the radiations in the far field over a certain bandwidth, the whole spatial spectrum is able to be retrieved. Before these operations, the system equation $G(k_x)$ of the imaging device should be investigated. The received spatial spectrum $H(k_x)$ can be thought as the convolution of the input $F(k_x)$ (the spatial spectrum of the targets) and the system equation $G(k_x)$

$$H(k_x) = H(k_x) * G(k_x), \tag{6}$$

where * denoted the convolution calculation. To simplify the theoretical deduction, the $G(k_x)$ is treated as the delta function $\delta(k_x)$ with specified amplitude and phase responses $\Delta A$ and $\Delta \theta$. We rewrite the equation (6) as

$$H(k_x) = F(k_x) \cdot \Delta A(k_x) \exp(i\Delta\theta(k_x)), \tag{7}$$

in which $\Delta\theta(k_x)$ is expressed by

$$\Delta\theta(k_x) = \int k_s(k_x)dx + \int k_{TEM}(k_x)dx + k_0 L, \tag{8}$$

where the first integral represents the phase accumulation by the SSP propagation along the uniform and graded parts, the second integral represents the phase change induced by the PPWG including both the uniform and tapered parts, and the third issue is caused by the transmission of radiated waves in the free space, where $L$ is the distance between the coupler and the receiving antenna. The amplitude response $\Delta A$ is induced by the imperfect coupling, propagating loss, reflection of couplers, and transmission loss between coupler and receiving antennas. Presenting a rigorous theoretical deduction of $\Delta A$ can be very difficult and time-wasting. Here, for the purpose of giving a principal verification of this imaging method, $\Delta A$ is thought as a constant for any value of $k_x$. Actually, $\Delta A$ and $\Delta \theta$ can be obtained by the pre-measurement, i.e., measuring the receiving signal $H(k_x)$, then $\Delta A(k_x)\exp(i\Delta\theta(k_x)) = H(k_x)/F(k_x)$.

To complete the imaging process, the imaging device is utilized to calculation. The length of uniform grating is 7.6 mm, and the distance between the coupler and the recording antenna is 1.5 mm. Other parameters are the same with above. The targets are localized at the center of this structure, 100 μm above the grating, avoiding the severe decay of evanescent waves before sampled by SSP structure. The phase retardation $\Delta\theta$ from $k_c$ to $4k_c$ is calculated through equation (8) and presented in Fig. 4(a). The dashed line in Fig. 4(a) is the simulated results by FEM, matching well with the theory. Obviously, $\Delta\theta$ varies quasi-periodically with the change of parallel wavevector, which consists with the theoretical prediction.

Then, we move step to extract the spatial spectrum of two slits, with 100 μm width and 200 μm (about $0.17\lambda$ for 0.26 THz) center-to-center distance. The calculation frequency range is from 0.16 to 0.26 THz. The amplitude profile of the retrieved spatial spectrum by our imaging device is rendered in Fig. 4(b). Compared with the theoretical result calculated by equation 1, we find this sampling device remarkably covering the spectral band $0.6k_c < |k_x| < 4.8k_c$, even within the deep evanescent domain. This provides a direct evidence of the superresolution imaging capability. Note that, the working spectrum can be further significantly enhanced by optimizing the parameters of the SSP device, since the maximum $k_x = \pi/p$. To visualize the field evolution in the sampling process, we illustrated the electric field distribution related to $k_x = 1.5k_c$ and $k_x = 2.9k_c$ in Fig. 4(c) and (d). Clearly, in the case of $k_x = 1.5k_c$, the evanescent component is efficiently converted into the propagating SSP, and then radiated into far-field with the assist of two couplers. While under the circumstance that $k_x = 2.9k_c$, the launched SSP from two slits will interference destructively with each other, resulting in no excited SSP.

The amplitude of the information of $|k_x| > 4.8k_c$ dramatically decreases, which is attributed to severe decay of deep evanescent component before reaching our device. This attenuation is able to be compensated by the pre-measurement. For the purpose of simply manifesting our imaging mechanism, the compensation is not taken into consideration in this work. Due to the limited working frequency band and deteriorating coupling efficiency, the spatial information of $|k_x| < 0.6k_c$ is lost. We can cover this part by a single shoot at far field, as they are all propagating waves. The spectrum of this part is depicted in Fig. 4(f), which is obtained by the spatial FT of the scattering far field at 0.26 THz. The retrieved phase is presented in Fig. 4(d), which match with the theoretical results. Compared with phase profile with Fig. 4(a), there are many fluctuations in Fig. 4(d). This is because the imperfect of two couplers. Part of SSP propagating back and forth along the grating induces reflections at both terminations, lending to Fabry-Perot-like resonance. Finally, the image is reconstructed by combing all the simulated results and performing IFT operations, as illustrated in Fig. 4(g). Notably, the objects are well distinguished, verifying the validity of the proposed far-field superresolution imaging method.

Constrained by the fabrication capabilities and test conditions, we experimentally demonstrate the imaging mechanism only in the microwave regime. This is totally acceptable as the proposed method is applicable at the longer wavelength, i.e., from to microwave to terahertz. And the analytical deduction is feasibility independent the working frequency. The chosen parameters are $p = 2$ mm, $d = 1$ mm, $L_1 = 45$ mm, $L_2 = 55$ mm, $L_3 = 50$ mm, $L_4 = 25$ mm, $L_5 = 8.2$ mm, $N = 9$, $\theta_1 = 14.6°$, $\theta_2 = 14.9°$, which is 20 times scaled from original simulation. The corresponding working frequency range is from 8 to 13 GHz. The image of the fabricated prototype is present in Fig. 5(a), along with two close shots of the uniform and gradient gratings. Two monopole antennas with 4 mm separation (0.17 λ with respect to 13 GHz), are used as the emitting sources. A horn antenna is employed to receive the radiated energy from the coupler. All the antennas are connected to a four-port vector network analyzer (Agilent N5245A PNA-X), as shown in the inset of Fig. 5(b). We only record the information of $-5k_c < k_x < -0.6k_c$ due to the total symmetrical structure. For the propagating part of $|k_x| < 0.6k_c$, it is obtained by recording the electric field in the far field at 13 GHz. By combing all the results, the retrieved phase and amplitude of the spatial spectrum are plotted in Fig. 5(b) and (c). The trend of experimental data satisfactorily consistent with the theory. Note that, compared with the theory, the measured data is slightly 'stretched'. The zero of the spatial spectrum appears in $k_x = 3.3k_c$ rather than $k_x = 2.9\ k_c$. The spectrum still owns considerable strength near $|k_x| = 5k_c$, while in the simulation only $|k_x| < 4.8k_c$ can be captured. The deviation results from the manufacturing tolerance, i.e., the depth of the groove is slightly less than 5.2 mm. This machining error leads to a blue shift of working frequency. From the reconstructed image in Fig 5(d), we can conclude that the two targets are clearly distinguished. The distance between two spikes in Fig. 5(d) is about 0.16λ, slightly smaller than the original 0.17λ, which is also attributed to the blue shift. By combing all the results, the capability of our imaging method can be fairly verified.

## 5. Two-dimension superresolution imaging

In the practical usage, two-dimensional (2D) subwavelength imaging ability is also desirable. All the discussion performed above only covers one-dimensional resolution, i.e., the thickness of SSP structure in the y direction is infinite and it has no response to the parallel wavevector $k_y$. To obtain 2D sub-diffraction-limited images, the information in all direction is needed. Therefore, the method of rotating imaging device in reference is employed, i.e., the imaging devices will be rotated 30 degree after each measurement, and then the spectral information of all directions can be retrieved. Firstly, we want to enhance the working performance of the imaging device. The parameters of the grating are optimized as $p = 120$ μm, $d = 60$ μm, $h = 255$ μm, the thickness of the grating in the y direction is truncated to 500 μm, other parameters keeps unchanged. The working frequency band is changed into 0.16 THz-0.274 THz. Under these parameters, the spatial spectrum of the grating can be extended to $8k_c$ (with respect to 0.274 THz). For a particular case, one 2D pattern is used to verified the 2D subwavelength imaging ability of our device. The electric field of 2D object is shown in Fig. 6(a), two $0.12\lambda$-sized squares with center-to-center distance of $0.25\lambda$. Its spatial spectrum is calculated and presented in Fig. 6(b). To reconstruct the spectrum, forgoing process of six measurements is employed. The propagating component of the spectrum is obtained by a single shoot at far field. The retrieved spectrum shown in Fig. 6(c) clearly reproduces the features of original data. Then, using IFT process, the 2D sub-diffraction resolution image is clearly distinguished, as illustrated in Fig. 6(d). Obviously, some details are lost when comparing the reconstructed image with Fig. 6(a). To further improve the quality of the imaging, one can increase the times of rotating measurement, with the price of time wasting. It is also feasible to replace the metallic grating with some 2D SSP structure [49], then the spectrum with all direction can be captured simultaneously.

**6. Discussion and conclusion**

In our demonstration, the simple method of frequency scanning is employed to tune the SSP mode. This brings two inevitably issues. Firstly, a terahertz source with the ability of tuning working frequency or providing broadband signal is needed. The time-domain terahertz spectroscopy or terahertz vector network analyzer can well fulfill this condition. Considering the cost, some diode oscillators are also available, for example, the products of Lytid. Secondly, the scattering fields of the targets may be changed through the imaging process, because the variable responses of the dielectrics at different frequencies. We can mitigate this situation by decreasing the frequency bandwidth. This is realized by optimizing the dispersion of the SSP structure, for example decreasing the duty ratio of the air in the subwavelength grating [29] or using TE-polarized SSP [25].

To complete the sampling process at a single frequency point, the SSP mode can also be changed using other manners. The first example is integrating the ultra-thin SSP structures with tunable lumped elements [43]. A simple design is presented in the inserted in the Fig. 7, where all the parameters are clearly labelled. Here, the varactor diode is used for tuning the wavevector of the SSP mode. Using the eigenmode method [50], the parallel wavevector of SSP at 10GHz as a function of capacitance is plotted in Fig. 7. As we can notice that the SSP mode with $k_0 < |k_s| < 8k_0$ can be consecutively excited by changing the capacitance of varactor diode, which is realized by tuning the bias voltage. Using the controllable dielectrics as the substrates or fillers of SSP structure are also feasible. For example, by adjusting the intensity of pump light shining on the silicon (Si) layer, the working properties of above

SSP mode are also tuned [42]. Therefore, the sampling process is quiet flexible, and superresolution imaging at a single frequency point is also available.

In conclusion, we propose and demonstrate that the far-field superresolution imaging by sampling the spatial spectrum in the microwave and terahertz frequency range. Both the theory and simulations indicate that the SSP structure behaves like a sensitive probe. The features of spatial harmonics can be well extracted by near-field SSP mode and sent into the far field, when they own the same parallel wavevectors. By tuning the parallel wavevector of the SSP mode, the broadband spatial spectrum of the targets is sampled and retrieved at far field. The image with resolution of 0.17 $\lambda$ is numerically and experimentally retrieved in almost real time. Most importantly, the imaging ability can be significantly enhanced by optimizing the parameters of the SSP structure. Our imaging mechanism also has ability on realizing 2D subwavelength imaging. Our work could find applications on non-destructive testing and manufacturing quality control.

**Acknowledgements**

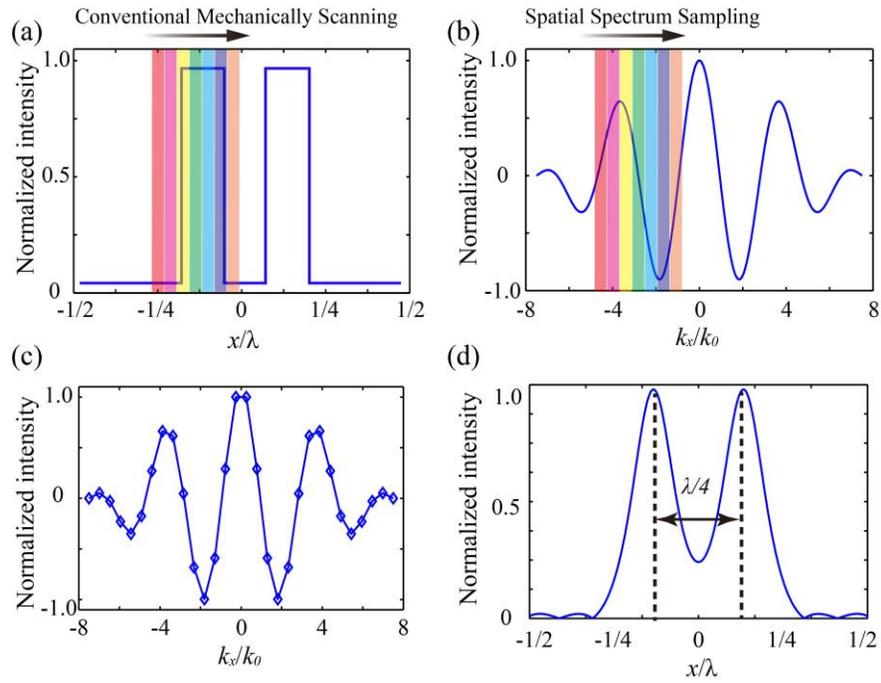

**Figure 1.** (a) and (b) present the imaging methods of conventionally mechanical scanning and our spatial spectrum sampling, respectively. (c) The extracted spatial spectrum after sampling process. (d) The reconstructed image by performing IFT process to the data in (c).

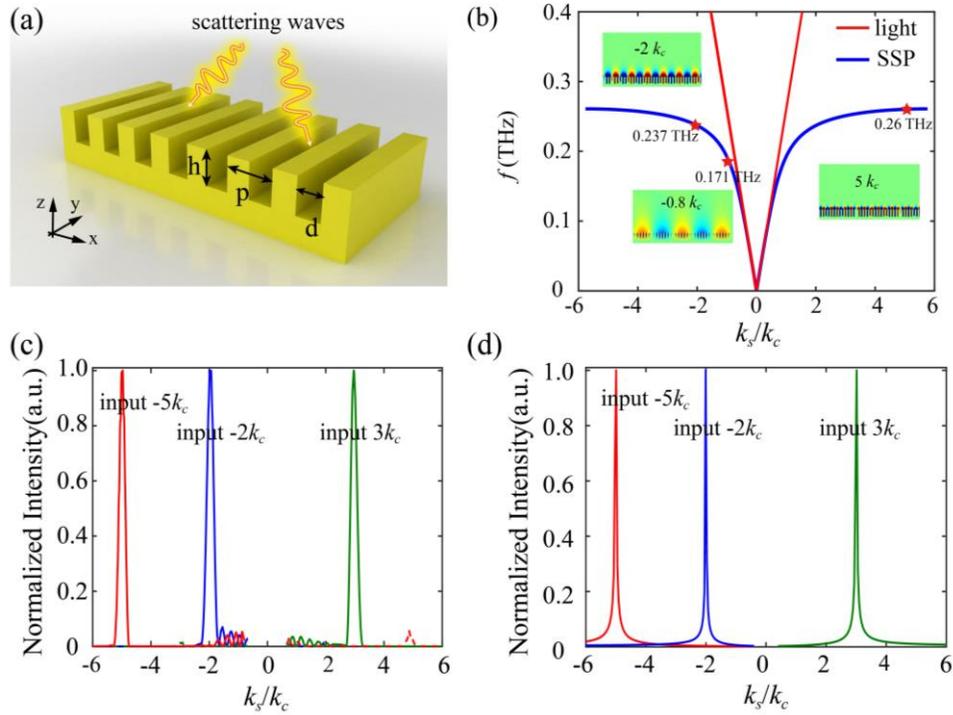

**Figure 2.** (a) the schematic of the subwavelength metallic grating; (b) The dispersive curve of the SSP supported by the gratings; the pictures in the inset are the electric field maps corresponding to $k_s = -0.8k_c$ (0.171 THz), $k_s = -2k_c$ (0.237 THz), $k_s = 5k_c$ (0.26 THz); (c) and (d) are the simulated and theoretical intensity of the excited SSP when the incident waves with parallel wavevector of $k_x = -5k_c$, $k_x = -2k_c$ and $k_x = 3k_c$.

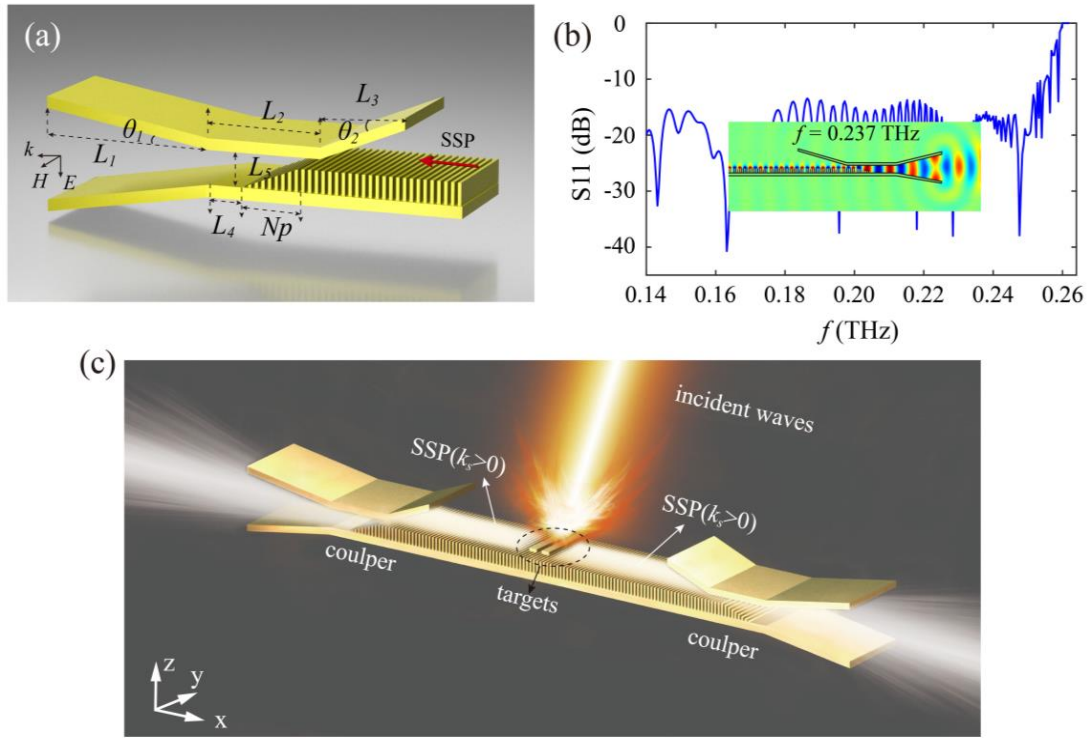

**Figure 3.** (a) the picture of SSP coupler, which consists of a tapered cover plate, a gradient grating part and a tapered PPWG; (b) the conversion coefficient between SSP and free space waves of this coupler. The inserted picture is the field evolution of the coupler when radiating SSP into space; (c) the working principle of our imaging device, including a uniform grating and two couplers.

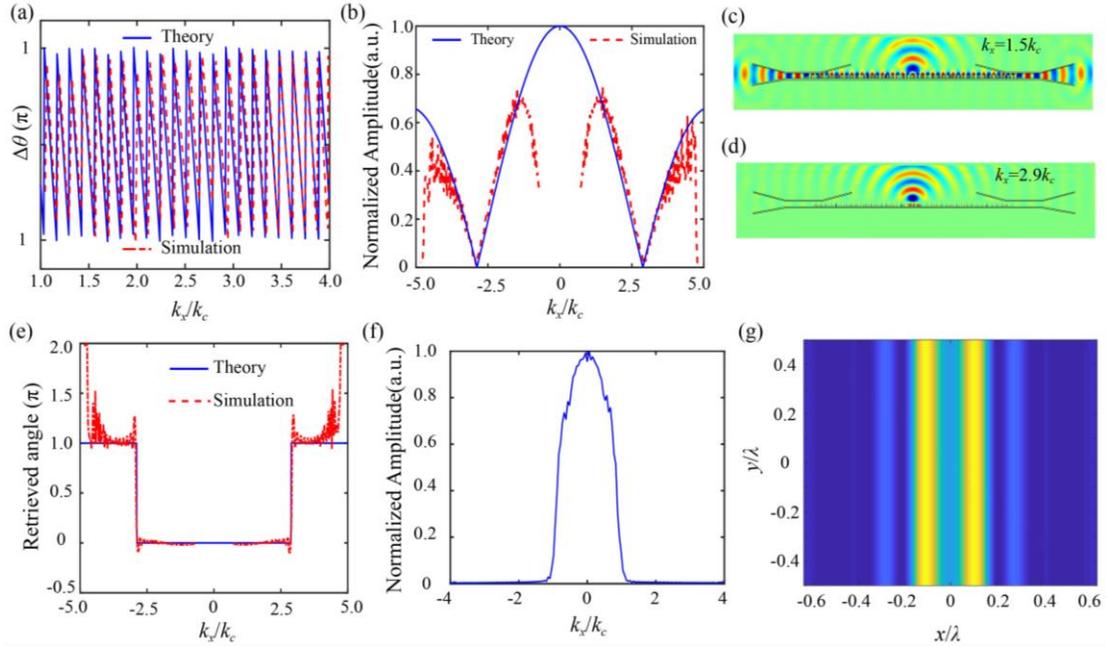

**Figure 4.** (a) the theoretical and numerical results of the phase retardation $\Delta\theta$ of this imaging device; (b) and (e) are the amplitude and phase profile of the spatial spectrum of the targets, respectively, where blue solid and red dashed lines represent the theoretical and retrieved results by our imaging device; (c) and (e) are the electric field distributions when the device sampling spatial harmonics with $k_x = 2k_c$ and $k_x = 3.5k_0$; (f) the retrieved spatial spectrum of propagating component by spatial FT process; (g) The reconstructed image by the IFT procedure, where two targets are clearly distinguished.

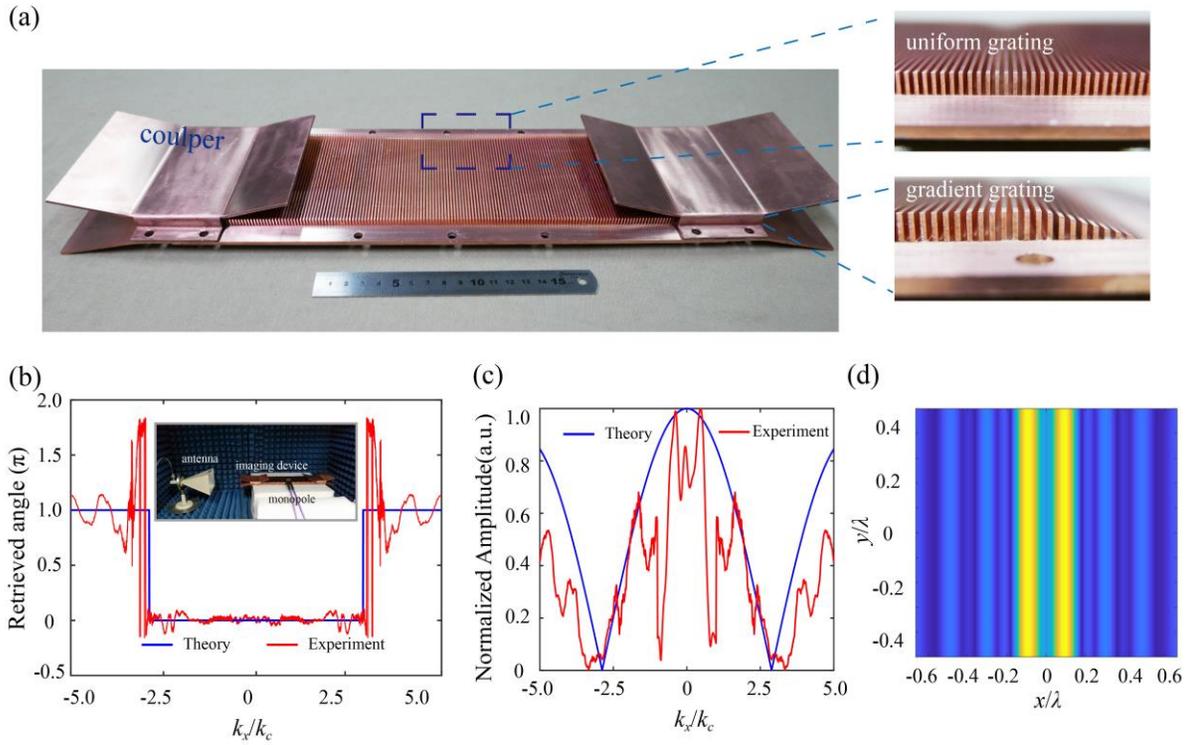

**Figure 5.** (a) the picture of the imaging device, companying with two close shot of uniform and gradient gratings; (b) and (c) the measured results of the phase and amplitude profiles of the targets. The inset of (b) is the picture of test scene; (d) shows the image constructed by the measured data.

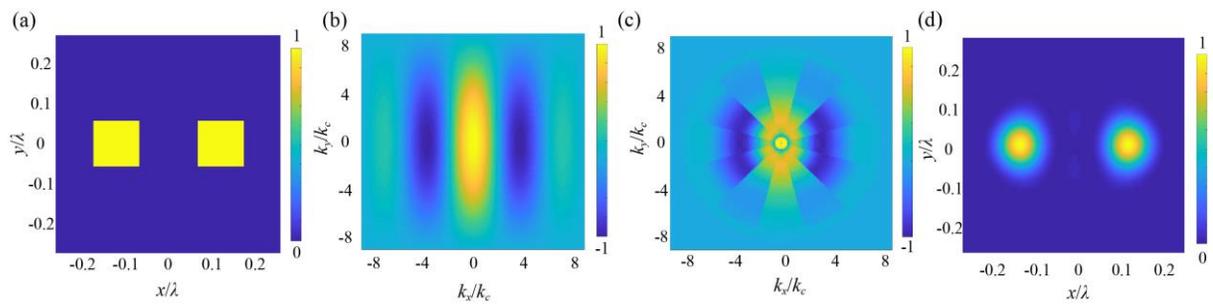

**Figure 6.** (a) and (b) are the electric field patterns of sub-diffraction object and its spatial spectrum. (c) the retrieved 2D spatial spectrum by rotating the SSP sampling device. (d) the reconstructed image using the spectrum present in (c).

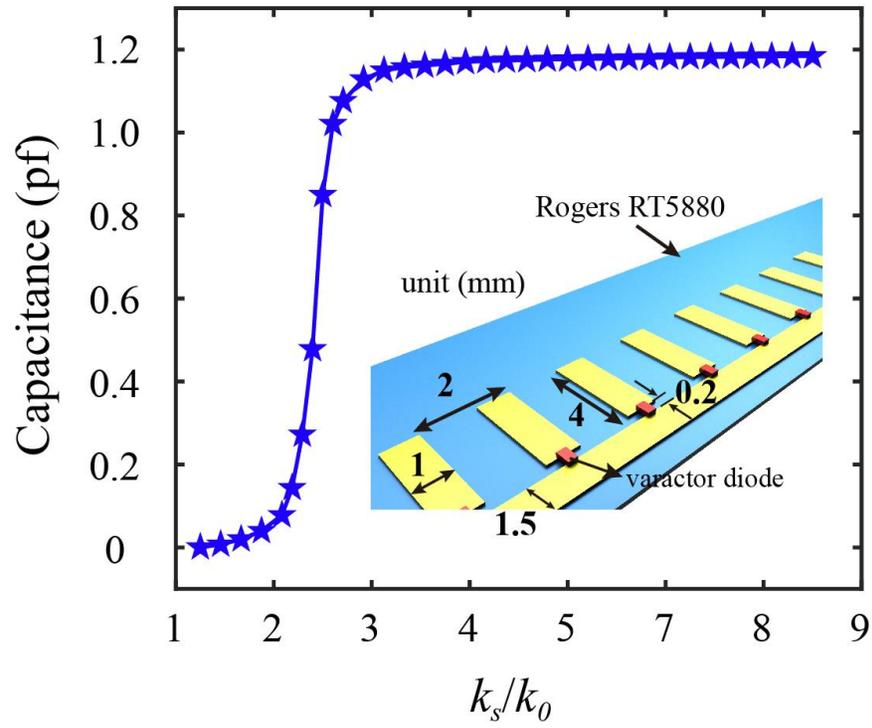

**Figure 7.** The parallel wavevector of the SSP mode as function of capacitance. The variation of capacitance can be realized by tuning varactor diode. The picture of the SSP structure is put in the inset, and the thickness of substrate and metal are respectively 0.018 mm and 1 mm.